\documentclass[twocolumn,twoside,dvips,showpacs,aps,pre,amsmath,amssymb]{revtex4}

\usepackage{graphicx}
\usepackage{dcolumn}
\usepackage{bm}
\usepackage{amsmath}
\usepackage{amssymb}
\usepackage{bbm}
\usepackage{mathrsfs}
\usepackage[sans]{dsfont}
\usepackage{epsfig,psfrag}
\usepackage{amsmath,amsfonts,amssymb}
\usepackage[usenames]{color}
\usepackage[dvips, bookmarks, colorlinks=true, plainpages = false, citecolor = blue, linkcolor = blue, urlcolor = blue, filecolor = blue]{hyperref}
\def\be{\begin{equation}}
\def\ee{\end{equation}}
\def\bea{\begin{eqnarray}}
\def\eea{\end{eqnarray}}

\newcommand{\ie}   {$\it{i.e.},~$}
\newcommand{\etal}   {$\it{et. al.},~$}

\begin{document}

\title{  Attracted Diffusion-Limited Aggregation }

\author{S. H. Ebrahimnazhad Rahbari$^{1,2}$}\email{sebrahi1@gwdg.de}
\author{ A. A. Saberi$^{3,4}$}\email{ab.saberi@ut.ac.ir}

\address { $^1$Plasma and Condensed Matter Computational Lab,
  Azarbayjan University of Tarbiat Moallam, Tabriz,
  Iran.\\ $^2$Max-Planck Institute for Dynamics and
  Self-Organization, Department of Complex Fluids, 37073 G\"ottingen,
  Germany\\ $^3$Department of Physics, University of Tehran, Post
  Office Box 14395-547, Tehran, Iran\\$^4$ Institut f\"ur Theoretische
  Physik, Universit\"at zu K\"oln, Z\"ulpicher Str. 77, 50937 K\"oln,
  Germany}

\date{\today}

\begin{abstract}
  In this paper, we present results of extensive Monte Carlo
  simulations of diffusion-limited aggregation (DLA) with a seed
  placed on an attractive plane as a simple model in connection with
  the electrical double layers. We compute the fractal dimension of
  the aggregated patterns as a function of the attraction strength
  $\alpha$. For the patterns grown in both two and three dimensions,
  the fractal dimension shows a significant dependence on the
  attraction strength for small values of $\alpha$, and approaches to
  that of the ordinary two-dimensional (2D) DLA in the limit of large
  $\alpha$. For non-attracting case with $\alpha=1$, our results in
  three dimensions reproduce the patterns of 3D ordinary DLA, while in
  two dimensions our model leads to formation of a compact cluster
  with dimension two.  For intermediate $\alpha$, the 3D clusters have
  \emph{quasi}-2D structure with a fractal dimension very close to
  that of the ordinary 2D-DLA. This allows one to control morphology
  of a growing cluster by tuning a single external parameter $\alpha$.
\end{abstract}


\maketitle

\section{INTRODUCTION}

By immersing an object with large surface-area-to-volume ratio into an
ionic solution, it is surrounded by a double layer of electrical
charge that significantly influences its surface
behavior~\cite{DL-1,DL-2}. Electrical attraction of free ions with
thermal motion in the fluid by the surface charge, forms a second
layer, known as diffuse layer, that shields out the Coulomb potential
of the surface layer and makes the whole structure electrically
neutral. Applications of double layer range from plasma
physics~\cite{chen}, to colloidal science~\cite{hunter}, and micro-
and nano-fluidics~\cite{kirby}.

As a simple theoretical model for the formation of a diffuse layer
in a special case, we implement extensive Monte Carlo simulations
for attracted random walks (ARW) introduced by Saberi~\cite{Saberi},
in order to study the formation of ionic aggregates on and near an
infinite plane located at $z=0$. An infinite surface charge exerts a
uniform constant electric force on the free ionic particles in the
fluid. It is, therefore, reasonable to consider an infinite
attractive plane of uniform attraction strength $\alpha$ which acts
on a free random walker launched from a point far from the
attractive plane. An ionic \emph{seed} is located at the center of
the infinite plane. Upon contacting, the free ionic random walker
sticks irreversibly to the cluster due to an absorbing bond
introduced between the ionic particles.

The diffusion of an attracted Brownian particle of mass $\emph{m}$ in
a fluid may be given by the following Langevin equation
\be\label{Eq1}m\ddot{z}+\gamma\dot{z}+\alpha\hspace{0.5mm}
\textrm{sgn}(z)=\xi(t),\ee where the second term denotes a
viscous-like friction force with drag $\gamma$, the third term stands
for an attraction force of strength $\alpha$ exerted by an infinite
plane at $z=0$, and $\xi(t)$ is a Gaussian white noise characterized
by \be\label{Eq2}\langle\xi(t)\rangle=0,\hspace{0.7cm}
\langle\xi(t)\xi(t')\rangle=2D\delta(t-t'),\ee with $D$ being the
diffusion coefficient. \\While such a Langevin-like equation
(\ref{Eq1}), to our best knowledge, has not been studied previously,
we implement such a process to produce a diffusion-limited aggregated
pattern on an attractive plane and investigate its fractal properties
in terms of the strength of attraction $\alpha$.  We apply two
different mechanisms in our model to produce the aggregates. First,
the aggregates are let to grow freely in three dimensions, \ie the
3D-ARW sticks to the aggregate upon touching the cluster giving rise
to the formation of a 3D pattern. Second, the aggregates are
restricted to only grow within the attractive plane, \ie the 3D-ARW
sticks to the aggregate only if it touches the cluster within the
attractive plane, which leads to the formation of a 2D pattern.\\ Our
results indicate that for small $\alpha$, the fractal dimensions for
both cases depend significantly on the strength of attraction, and in
the limit of large $\alpha$, they converge to the fractal dimension of
the ordinary two-dimensional (2D) diffusion-limited aggregation
(DLA). This rapid convergence, however, implies the formation of a
\emph{quasi}-2D pattern for a 3D aggregate for intermediate vales of
$\alpha$. Furthermore, for the two limiting cases, \ie $\alpha=1$ and
$\alpha\rightarrow\infty$, our model reproduces the results of the
ordinary DLA. Loosely speaking, for $\alpha=1$, the 3D-ARW is
identical to the ordinary 3D-RW and thus our model generates the same
patterns as the ordinary DLA in three dimensions.  In the limit of
$\alpha\rightarrow\infty$, the 3D-ARW can only move within the
attractive plane and is indeed identical to the 2D-RW and thus, in
both cases, the patterns of the ordinary DLA in two dimensions are
reproduced. The only difference is for $\alpha=1$ in two dimensions,
where our model gives a new class of aggregates which are compact
clusters with dimension two.

The ordinary DLA, known also as a Brownian tree, was originally
introduced by Witten and Sander \cite{Ref1} to model the aggregates of
metal particles formed by adhesive contact in low concentration
limit. This model is shown to describe many pattern formation
processes including dielectric breakdown \cite{Ref2}, electrochemical
deposition \cite{Ref3, Ref4}, viscous fingering and Laplacian growth
\cite{Ref5}. According to this model, the walker does a random walk
process initiated from infinity like as described above in our model
but without experiencing any external force. Many of its fractal
properties have then been reported from various simulations done both
in real and mathematical plane (see for example \cite{DLA} and
references therein).

This study uncovers a new insight into various related phenomena with
a discrete time lattice walk \cite{appl0, appl1}, relaxation phenomena
\cite{relax}, exciton trapping \cite{trap} and diffusion-limited
reactions \cite{appl1, react}.

\begin{figure}[b]
  \includegraphics[width=0.43\textwidth]{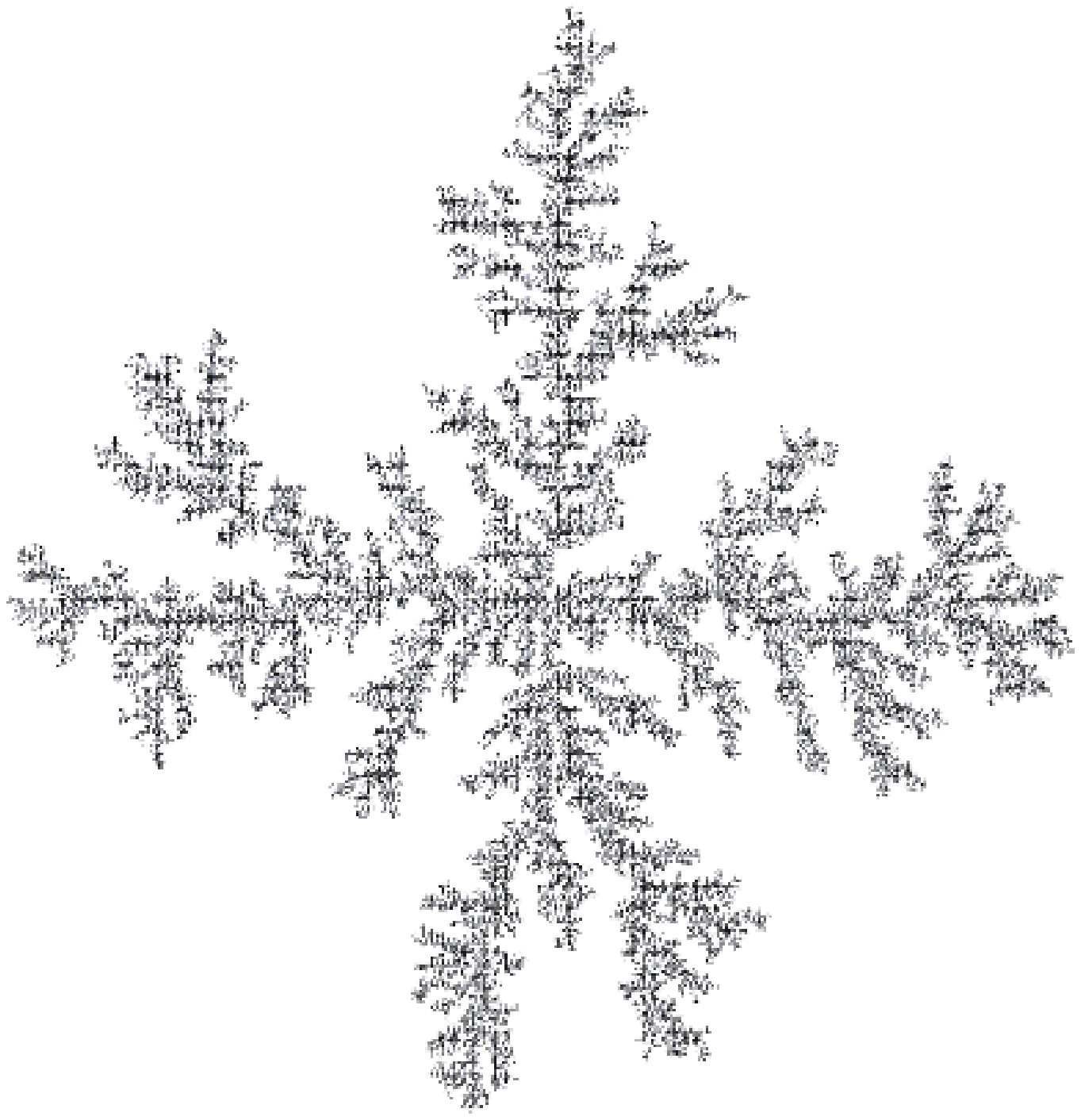}
  \includegraphics[width=0.43\textwidth]{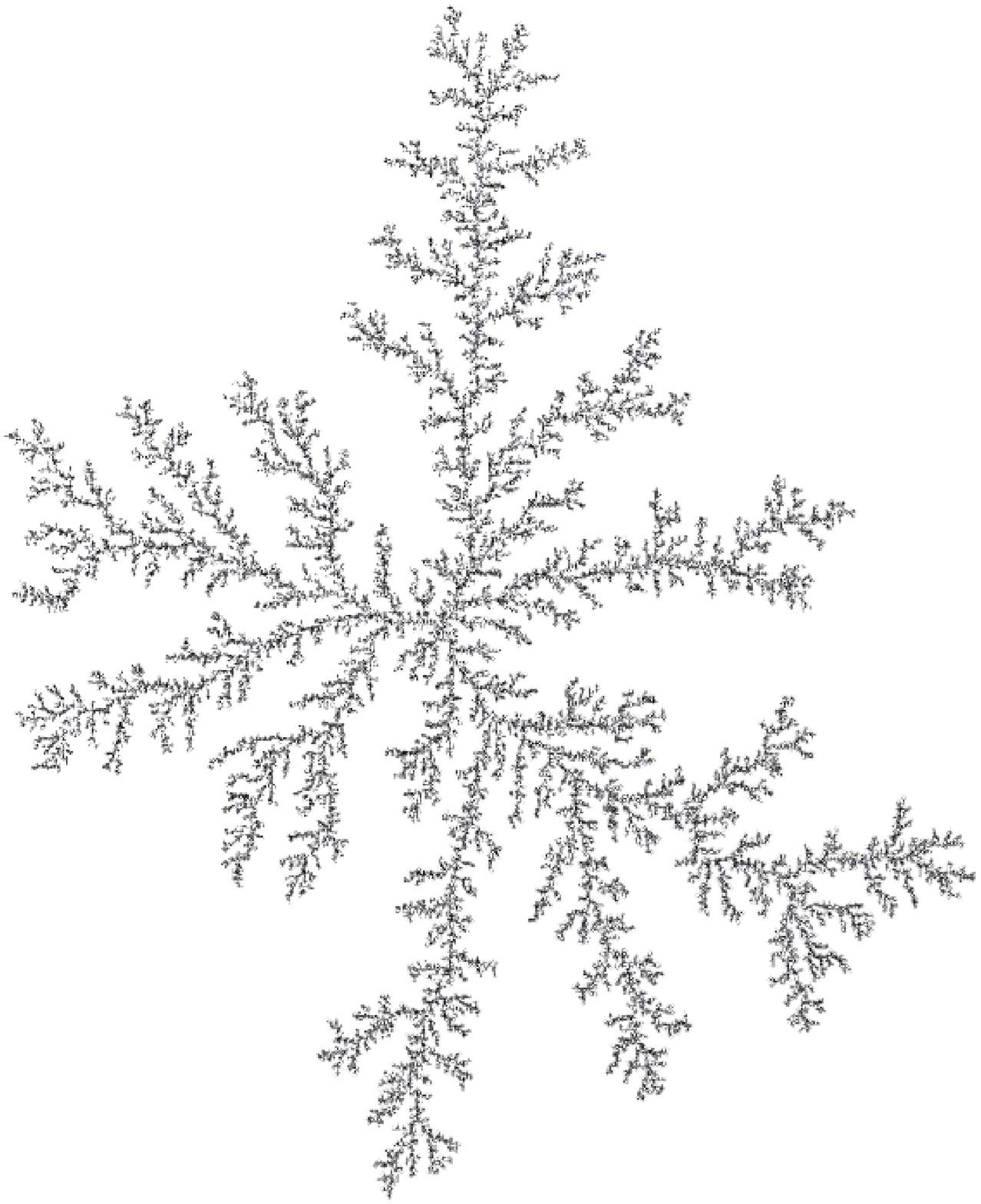}
  \caption{Snapshots of the growing clusters for $\alpha = 1.2$ (Top),
    and $\alpha = 10$ (Bottom), with the same number of particles $M =
    10^5$. The radius of gyration is $R_G = 251.5$ and $R_G = 410.54$
    for the top and bottom figures, respectively. Distance from camera
    is the same for both clusters. One sees that the gyration radius
    of the clusters, $R_G$, increases with the attraction strength $\alpha$.
    Moreover, for small $\alpha$ the cluster has fat arms, while for larger $\alpha$ the chains are
    much elongated and delicate. The pictures are rendered by the {\em
      Povray} software.
    \label{fig:clusters_2D}}
\end{figure}

\section{Attracted diffusion-limited aggregation (ADLA)$-$details of simulation}
\label{sec:DLA}

In this section, we investigate aggregation of ARWs on and near an
attractive plane based on the algorithm proposed in \cite{Ref1} for
ordinary DLA clusters, accompanied by the details of simulations.  For
$M$ being the total number of particles, or equivalently the cluster
\emph{mass}, the simulation box is considered to be a cube of size
$L_x = L_y = 20 \sqrt{M}$ and $L_z = 60 \sqrt{M}$. The box is divided
into a regular lattice of $L_x \times L_y \times L_z$ cubic cells of
unit size. Periodic boundary conditions are applied along all three
directions. The attractive plane is a horizontal plane at
$z=\frac{L_z}{2}:=0$ whose bottom-left corner is taken as origin of
the coordinate system. We store information of each lattice site by an
integer in the computer. Accordingly, the total memory to be occupied
by the lattice will be equal to $4L_xL_yLz \approx 10^5 M ^{3/2}$
Bytes. One sees that as the total number of particles $M$ increases,
the total memory grows rapidly and for large $M$ the total memory
becomes so huge that can not be handled by conventional computers. In
order to prevent excessive memory usage, we apply different
restrictions to our algorithm to reduce the memory. These tricks will
separately be discussed for both $2D$ and $3D$ clusters in
Sec.~\ref{sec:2D} and Sec.~\ref{sec:3D} respectively.

First, let us briefly review the probabilities governing the movement
of an ARW~\cite{Saberi}. For the random walker at $z\neq 0$, there are
six directions which can be chosen by the walker. One is that points
towards the plane whose probability is set to $\alpha p$, where
$\alpha > 1$ is strength of the attraction. For the five directions
being left, the corresponding probability of each link is equally set
to $p$, providing $ p = \frac{1}{\alpha + 5}$. For the random walker
at $z = 0$, there are two directions perpendicular to the plane whose
probability is $p^\prime$, and four directions on the plane each of
which with the probability $\alpha p^\prime$, such that $p^\prime =
\frac{1}{4\alpha + 2}$. This setting of probabilities guaranties that
the walker likes to ramble on and around the attractive plane.

It is well known that formation of DLAs is dominated by diffusive
motion rather than the convective one. Therefore, the walker must
start to move far enough away from the attractive plane. Once a seed
particle is settled on the attractive plane, a random walker is
introduced on a random lateral position at height $z_0 = 15 \sqrt{M}$,
and moves according to the probabilities having been described
above. The walker keeps moving until it sticks to the seed. Following
that, a new random walker is introduced at $z_0 = 15 \sqrt{M}$ from a
random lateral position, and is let to move until it sticks to one of
those particles being frozen. This procedure is applied until the
desired number of the particles in the cluster, \ie $M$, is
reached. For the growth of 2D and 3D structures, we implement
different additional rules that will be discussed in Sec.~\ref{sec:2D}
and Sec.~\ref{sec:3D}.

\subsection{2D structures}
\label{sec:2D}

2D structures are formed based on a {\em land-escape} procedure.  When
a particle is either at $z = +1$ or $z = -1$, it may {\em land} on the
plane if its shadow on the plane is not occupied, otherwise it must
{\em escape} to one of the five remaining directions with equal
probabilities. Once the particle has landed on the plane, it will
diffuse according to the aforementioned rules until it finds at least
one frozen particle on its neighborhood and gets stuck to that
point. It should be stressed out that according to the probabilities
governing the motion of the ARWs, there still exists a probability for
the walker to break from the plane and fly back into the space.

From our numerical experiments, we empirically found that the radial
distance of the farthest particles from the seed never exceeds $r =
8\sqrt{M}$ limit. We used this fact to accelerate our simulation runs,
in a way that the attractive plane is being divided into two distinct
regions, the inner region where the cluster forms, and, the outer
region where the random walker moves freely with no need for exploring
a possible occupied nearest-neighbor site. We employed this fact in
our simulations and found that it considerably decreases the
computational effort.

In figure~\ref{fig:clusters_2D}, two snapshots of the growing
clusters on the attractive plane are shown for two different
strength of attraction $\alpha = 1.2$ and $\alpha = 10$, shown at
the top and bottom of the figure, respectively. The cluster mass $M
= 10^5$ and the distance from camera are considered to be the same
for both patterns for a comparison.\\Both figures have a fractal
structure with a rotational symmetry about the \emph{z}-axis and the
same number of main branches. For lower $\alpha$, the particles are
being condensed along the main arms giving rise to much fatter
branches than for the larger $\alpha$ for which, the cluster has
delicate chains elongated on the attractive plane with deeper fjords
and sharper tips. It is also evident that the gyration radius $R_G$
of the clusters increases with the attraction strength $\alpha$.
Here, the gyration radius $-$in units of a particle diameter$-$ is
found to be $R_G = 220.8$ and $R_G = 410.54$ for $\alpha = 1.2$ and
$\alpha = 10$, respectively.

\begin{figure}[t]
  \[
  \includegraphics[width=0.45\textwidth]{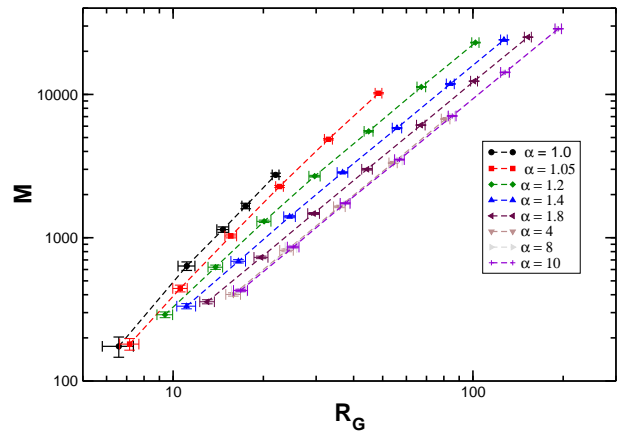}
  \]
  \caption{Cluster mass $M$ as a function of the radius of gyration
    $R_G$ of the 2D growing clusters on the attractive plane for
    different values of $\alpha$. For $\alpha\gtrsim 1.15$, the curves
    exhibit a full power-law behavior Eq. (\ref{eq:M_Rg}).
    \label{fig:N_Rg_2D}}
\end{figure}

\begin{figure}[h]
  \[
  \includegraphics[width=0.45\textwidth]{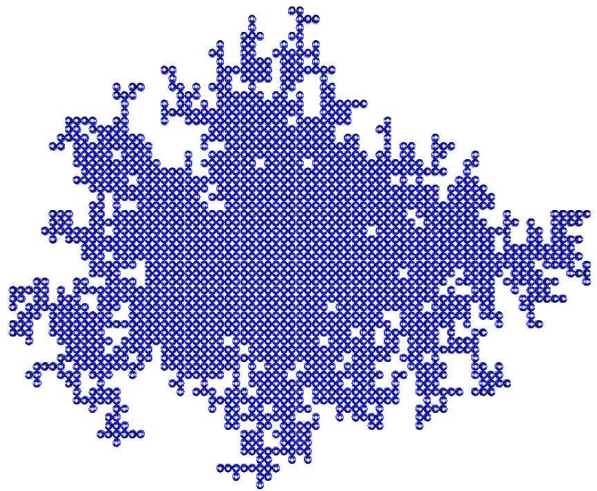}
  \]
  \caption{A typical example of a 2D cluster grown on the attractive plane with
  $\alpha=1$. The number of particles is $M=1875$ which forms a compact cluster of
  dimension $\approx$2, with gyration radius $Rg\simeq19.62$.
    \label{Fig3}}
\end{figure}

In order to get a more quantitative insight about the structure of the
clusters, we compute the gyration radius $R_G$ of different clusters
grown under the same conditions but with different strength of
attraction $\alpha$. Figure \ref{fig:N_Rg_2D} illustrates the cluster
mass $M$ as a function of the gyration radius $R_G$ of the clusters
for various $\alpha$. Since the ensemble averaging plays a crucial
role in such computations, the averages, for each point in the
Fig.~\ref{fig:N_Rg_2D}, are taken over $90$ different clusters for
given values of $\alpha$ and $M$. It is well-known that the following
relation is usually held for fractal patterns
\begin{equation}
M \propto R_G^{D_f}, \label{eq:M_Rg}
\end{equation}
where $D_f$ is the fractal dimension of the cluster. Our results are
in accord with this power-law relation (\ref{eq:M_Rg}) for
$\alpha\gtrsim 1.15$. For smaller values of $\alpha$, the ARW crosses
over to the ordinary 3D-RW and starts to fill in the empty regions
between the branches. This leads the cluster to being much
condensed. A typical example of such a compact cluster for $\alpha=1$
is shown in Fig.~\ref{Fig3}. Although the power-law behavior is
disturbed, within our numerical accuracy, for $\alpha\rightarrow1$,
their outer perimeter may keep their fractal properties as it comes
out from Fig.~\ref{Fig3} \footnote{Similar to the compact islands
  appear in a 2D cross-section of a height profile in a
  (2+1)-dimensional Kardar-Parisi-Zhang equation \cite{KPZ}.}.  This
would be investigated in more detail in a future work.

According to our model, for $\alpha = 1$ the attraction strength is
set to zero. As a principle of DLA, random walkers should initiate
their motion from the infinity (far from the seed). Therefore, there
is a very little chance to find a random walker, triggered at
infinity, on the surface of the attractive plane. Within this small
chance, again there is a very little probability to keep the
particle moving on and around the plane. Once the particle lands on
the plane, normally it tends to escapes to infinity and hardly finds
a frozen particle in its neighborhood to join to the cluster.
Consequently, for 2D clusters at $\alpha = 1$ the growth rate is
almost zero and we observe only clusters with small mass $500 < M <
3000$ which is not enough to examine any scaling relation.
Therefore, we get a poor statistics for $\alpha = 1$ (see also
Fig.\ref{fig:2D_3D_compare}).

\subsection{3D structures}
\label{sec:3D}

In contrast to 2D structures, the growth model for 3D clusters is not
restricted to the surface of the attractive plane. Instead, the random
walker in this case can get stuck in any lattice site whose each of
the nearest-neighbor sites is already occupied, whether or not being
in the attractive plane.

\begin{figure}[h]
  \includegraphics[width=0.4\textwidth]{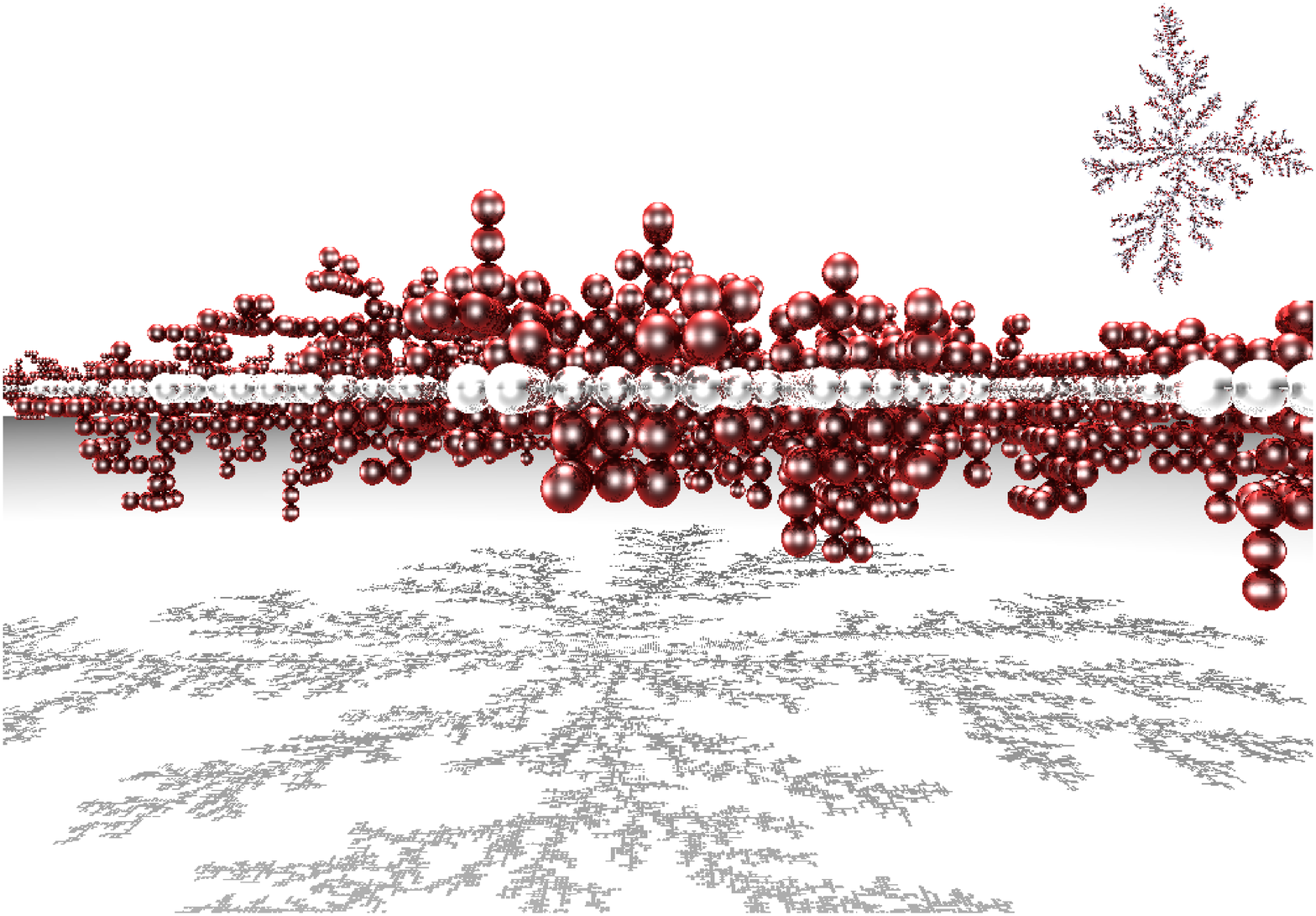}
  \includegraphics[width=0.4\textwidth]{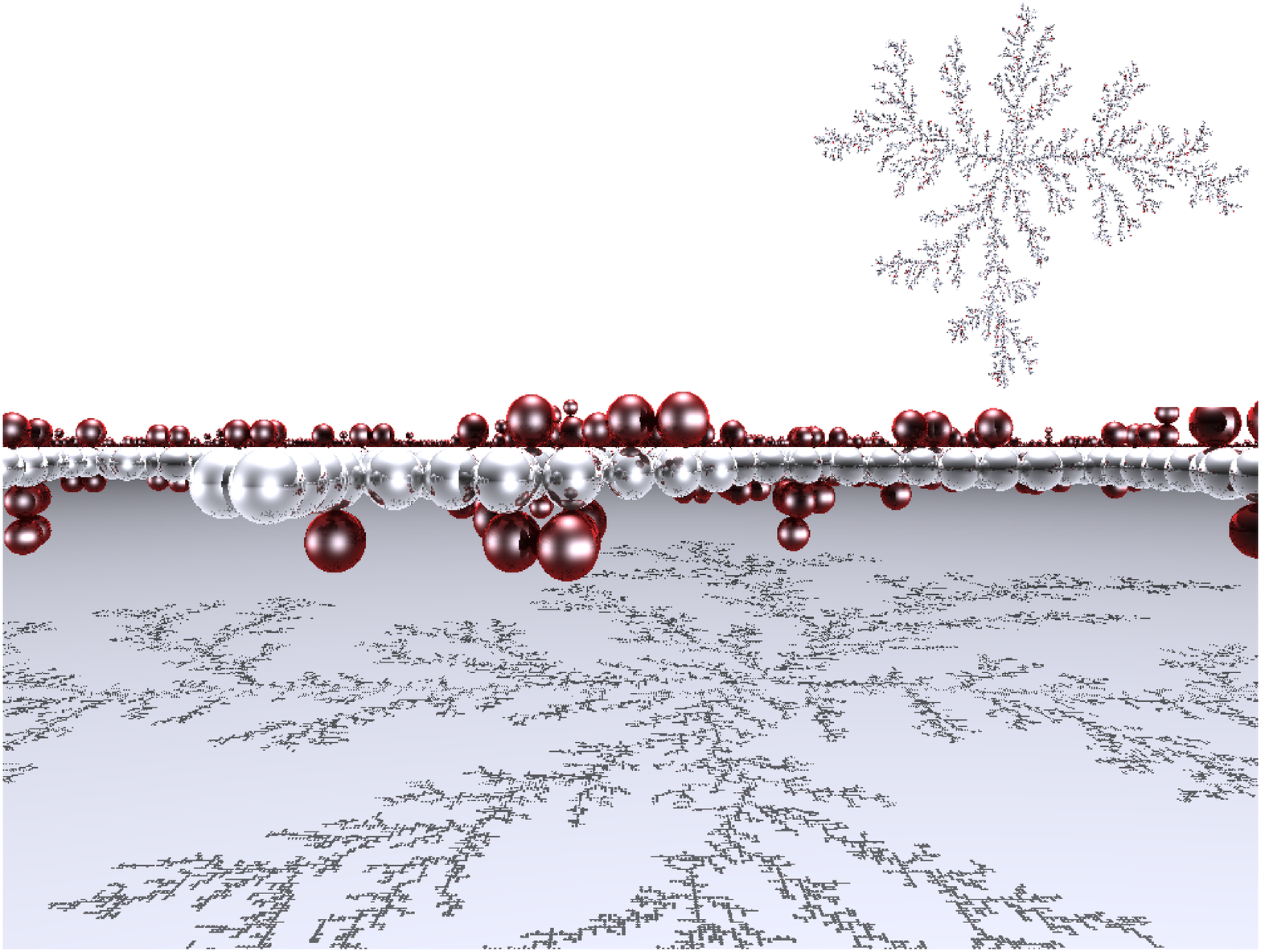}
  \includegraphics[width=0.4\textwidth]{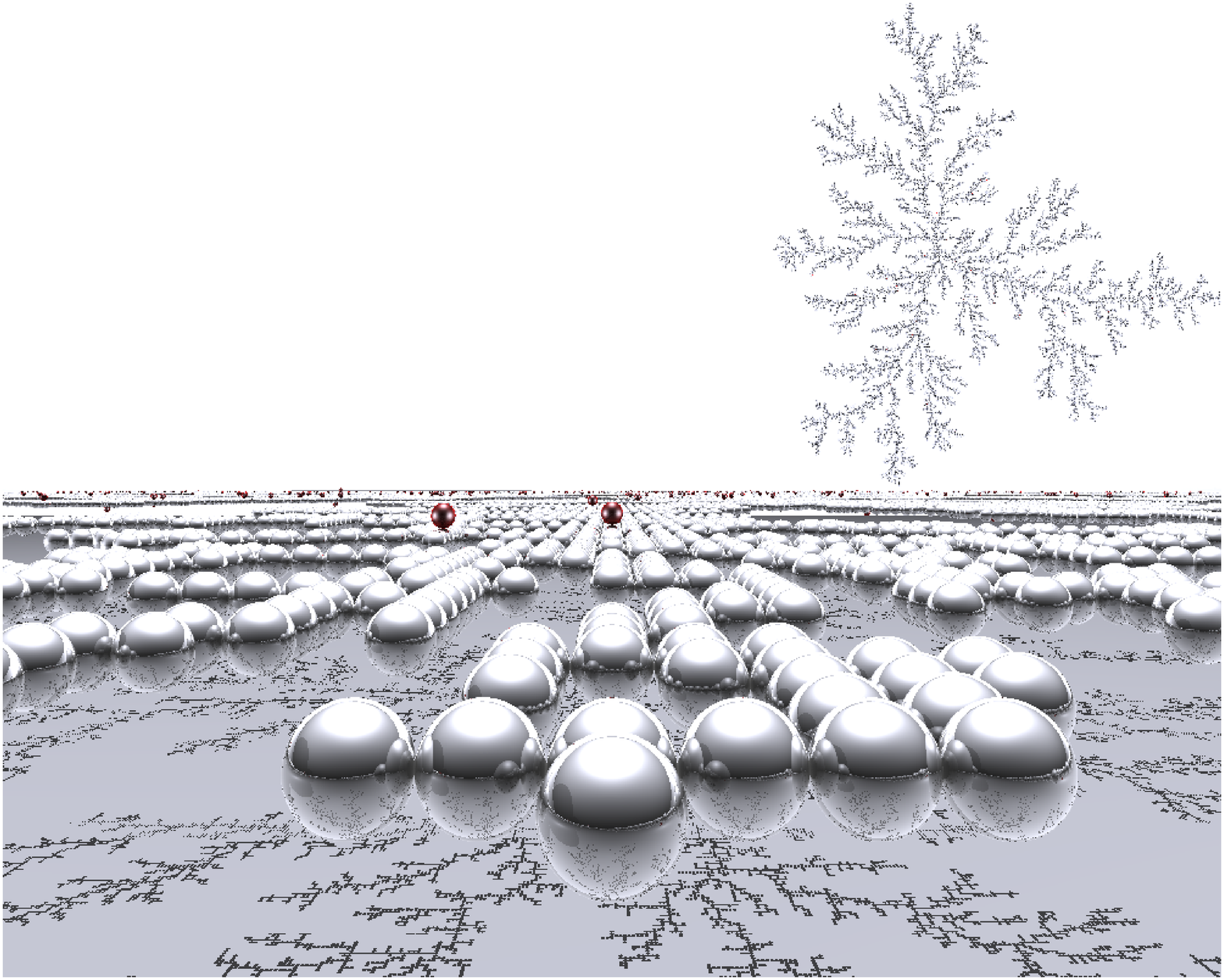}
  \caption{Three typical examples of 3D grown clusters of number of
    particles $M = 10^5$. The metallic (gray) particles lie on the
    attractive plane, and the rest$-$not on the surface$-$are
    illustrated by the red (dark) color. The strength of attraction is
    considered to be $\alpha = 1.2, 2$ and $10$, with $R_G = 199.6,
    342.8$ and $405.9$, from top to bottom, respectively. The maximum
    height of the towers perpendicular to the attractive plane never
    exceeds 10 particle diameter on either side of the attractive
    plane, implying the formation of \emph{quasi}-2D clusters.
    \label{fig:3D_snapshot}}
\end{figure}

In figure~\ref{fig:3D_snapshot}, we show some of the resulting
patterns obtained by this algorithm in three dimensions. The figure
shows a side view of the grown clusters consisting $M = 10^5$
particles with $\alpha =1.2, 2$ and $10$, with $R_G = 199.6, 342.8$
and $405.9$, from top to bottom, respectively. Two different colors
are chosen for the particles: the metallic (gray) color for those
sitting on the attractive plane at $z = 0$, and the red (dark) color
for particles sitting elsewhere with $z \neq 0$. As can be seen from
the figure, the number of red (dark) particles is significantly
dependent on the strength of attraction. For small $\alpha$, the ARW
can easily wander around and thus, it is more likely for the particle
to get a chance to stick to the cluster somewhere at $z \neq 0$, while
for larger $\alpha$, the ARW is almost rambling on and near the
attractive plane which increases the sticking probability within the
plane. In the limit of $\alpha\rightarrow \infty$, the 3D-ARW falls
onto the ordinary 2D-RW giving rise to the formation of ordinary DLA
clusters in pure two dimensions. For $\alpha=1$, at the other limiting
side, the 3D-ARW is the same as the ordinary 3D-RW and we get back the
ordinary 3D-DLA patterns.

We empirically found that for $1 < \alpha < 1.2$, the thickness of 3D
clusters on either side of the attractive plane never exceeds $35$
particle diameter, even for clusters of size up to $M=10^5$. Also, for
$\alpha > 1.2$, the height of the tallest tower in the cluster never
reaches $10$. Based on this experimental evidence, similar to that
of the 2D case, the simulation box can technically be divided into two
following distinct volumes to reduce the memory. An inner volume \ie
with $-35\leq z\leq35$ for $1 < \alpha < 1.2$, and $-10\leq z\leq10$ for
$\alpha > 1.2$, where the cluster can form and the rest as the outer
volume where the particle moves freely without an extra check for
looking up for an occupied nearest-neighbor site. We applied this fact
in the structure of our programming codes and found that it
considerably speeds up the simulations.

\begin{figure}
  \[
  \includegraphics[width=0.45\textwidth]{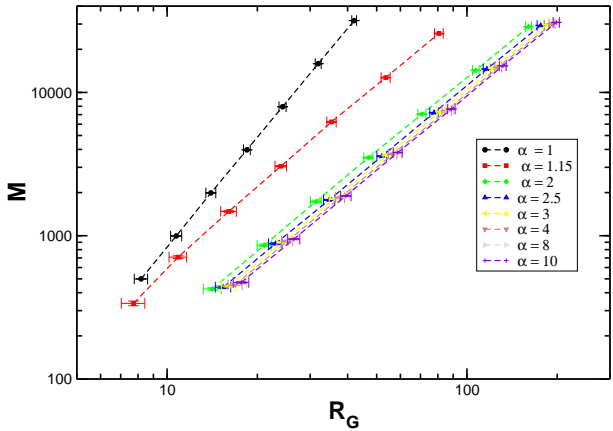}
  \]
  \caption{Cluster mass $M$ as a function of the radius of gyration
  $R_G$ of the 3D growing clusters formed around the seed placed on the attractive plane
  for different values of $\alpha$.
    \label{fig:N_Rg_3D}}
\end{figure}

The quantity of interest is again the gyration radius, $R_G$, for
clusters of different mass $M$. Figure~\ref{fig:N_Rg_3D} shows our
results of cluster mass for 3D clusters as a function of their
gyration radius for different attraction strength $\alpha$. Each
point in Fig.~\ref{fig:N_Rg_3D} is averaged over an ensemble of $90$
independently grown samples. The mass of clusters range between $500
\le M \le 32000$. Error-bars for both $R_G$ and $M$ are visible in
the plot. We found a perfect scaling behavior for $\alpha \ge 2$,
whereas for $1< \alpha < 2$ our data fits to a power-law with higher
uncertainty (this is also much evident from the inset of Fig.
\ref{fig:2D_3D_compare}). One also sees that for a constant cluster
mass, the gyration radius increases upon increasing the attraction
strength which is due to the elongation of the growing cluster
within the attractive plane. As mentioned above, the thickness $d$
of the clusters along the \emph{z}-axis never exceeds $70$ particle
diameter even for clusters of size up to $M=10^5$. This implies a
considerably small relative thickness, defined as $d/R_G\ll1$, for
the patterns, and thus indication of the formation of
\emph{quasi}-2D clusters. This would later be confirmed in the next
section where we compute the fractal dimension of the grown
patterns.

\section{Fractal dimension of 2D and 3D ADLA clusters}

In this short section, we compute the fractal dimension of the
clusters as a measure characterizing the statistical complexity of
the grown fractal patterns in both two and three dimensions. The
fractal dimension of the clusters can easily be calculated from the
data given in Fig.~\ref{fig:N_Rg_2D} and Fig.~\ref{fig:N_Rg_3D} by
examining the scaling relation (\ref{eq:M_Rg}).

\begin{figure}
  \[
  \includegraphics[width=0.45\textwidth]{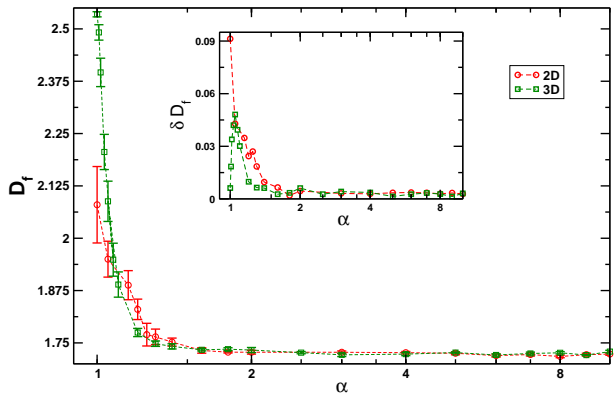}
  \]
   \caption{The fractal dimension $D_f$ of the attracted
    diffusion-limited aggregation clusters grown in two and three
    dimensions, represented by open circles ($\bigcirc$) and squares
    ($\square$), respectively, as a function of the attraction
    strength $\alpha$. The inset shows the corresponding error
    bars within which the power-law behavior Eq. (\ref{eq:M_Rg}) holds.
    \label{fig:2D_3D_compare}}
\end{figure}

In figure~\ref{fig:2D_3D_compare}, we compile our data for the
fractal dimensions of 2D and 3D clusters which are represented by
the open circles ($\bigcirc$) and squares ($\square$),
respectively.\\ For both 2D and 3D clusters, the fractal dimension
significantly depends on the strength of attraction $\alpha$,
especially for $\alpha<2$. The smaller the $\alpha$, the denser the
clusters are. For larger $\alpha$, the fractal dimensions in both
cases, are less dependent on $\alpha$, rapidly converging to the
value $D_f \approx 1.72$ which, within the statistical errors, is
almost in accord with that of the ordinary DLA clusters in two
dimensions with $D_f\approx1.71$.

As discussed before, the overall intermediate fractal behavior seems
to be dominated by the two crossover limits in the statistical
behavior of the underlying diffusion process \ie the crossover of
the 3D-ARW to ordinary 3D-RW and 2D-RW for $\alpha=1$ and
$\alpha\rightarrow \infty$, respectively.\\For $\alpha=1$, the 2D
clusters have a compact structure of dimension $\approx$ 2, and for
the 3D clusters we obtain a fractal dimension $D_f\approx2.53$ very
close to the expected value for the ordinary 3D-DLA \cite{3D-DLA}.

\section{Discussion and Conclusion}

In conclusion, we introduced a model of aggregation based on an
extension of diffusion-limited aggregation where the underlying
diffusion process is a 3D Brownian motion (or, a random walk on the
lattice) which is attracted by a plane with strength $\alpha$. The
\emph{seed} particle is placed on the attractive plane.\\In two
dimensions, the fractal properties of the aggregated cluster is
shown to be dependent on the attraction strength giving rise to
formation of patterns with $1.71 \lesssim D_f \leq 2$. The two
limiting values for the fractal dimension $D_f$, in the scaling
region, are discussed to be governed by the crossover of the 3D-ARW
to the ordinary 3D-RW for $\alpha\rightarrow1$ from one side, and
its convergence to the ordinary 2D-RW at the other side with
$\alpha\rightarrow\infty$, which leads to ordinary 2D-DLA patterns
of $D_f\simeq1.71$.

In three dimensions, the rotational symmetry, which is present in
the ordinary 3D-DLA clusters with $\alpha=1$, is broken by the
attractive plane with $\alpha>1$, and the model leads to formation
of clusters with $1.71 \lesssim D_f \lesssim 2.53$. For intermediate
$\alpha$, we obtain \emph{quasi}-2D clusters whose relative
thickness, defined as the thickness of a cluster perpendicular to
the attractive plane rescaled by its radius of gyration, is
relatively small. Our results indicate a scaling region with
$\alpha\gtrsim2$, in which the fractal structure of the clusters can
be characterized by a single fractal dimension $D_f\approx1.72$,
independent of $\alpha$.

Recently, there has been a growing attention and frequent reports on
the fact that morphology of self-assemblies can be controlled by
tuning various physical parameters such as the packing fraction
$\phi$, and the attraction strength. This is specifically important
for handling of soft materials and food processing. In our study, we
found out that a crossover from ramified, fractal clusters to
compact aggregates occur upon decreasing the attraction strength
$\alpha$ at the limit of small attraction strength. For the two
dimensional clusters the threshold is found to be $\alpha_c = 1.15$.
As a result, in our model morphologies of the growing clusters can
be controlled by tuning the attraction strength $\alpha$.

In a recent study of aggregation of hard
spheres~\cite{Poon_scaling_fail}, a similar crossover from a compact
to ramified aggregation is found which in the former regime the
scaling relation fails while in the latter case the scaling holds.
The crossover from compact to fractal clusters happens upon
increasing the packing fraction $\phi$. The threshold is found to be
at $\phi = 0.55$ which is very close to the glass transition point
at $\phi_g = 0.58$. Therefore, in this case, the structure of
aggregations can be controlled by the packing fraction $\phi$. Zhang
\etal~\cite{Zhang_scaling_fail} report results of experiments on
colloidal suspensions with short-range attraction and long-range
repulsion. They show that the morphology of the clusters can be
controlled by tuning either the attraction strength or the packing
fraction $\phi$ resulting to an elongated-to-branched crossover.

These reports imply that the compact-to-ramified crossover may be
considered as a universal property of aggregation of particles.
Consequently, this can suggest the possibility of proposing a single
phase-diagram that captures such behavior for different systems with
different underlying physical processes and inter-particle
interactions.

A possible future work for constructing a much more realistic model
of diffusive layer may be an extension of the present model by
considering a number of mobile seed particles on the attractive
plane that the growth of each can be given by our implemented
procedure. In that case, the growing clusters can stick together
upon contacting. The results will appear elsewhere in the near
future.\\

\section{Acknowledgments}

We would like to thank M. Ghorbanalilu for his fruitful comments.
SHER is partially supported with a fund number 218-D-3390, granted
by Azarbayjan university of Tarbiat Moallem. AAS acknowledges the
partial financial support by the research council of the University
of Tehran, and INSF grant as well.

\end{document}